\documentstyle[amssymb,twocolumn,prl,aps]{revtex}
\input epsf.sty
\newcommand{\ket}[1]{{|#1\rangle}}
\newcommand{\bra}[1]{{\langle#1|}}

\newcommand{\Tr}{\mathop{\rm Tr}\nolimits}
\newcommand{\wt}{\mathop{\rm wt}\nolimits}

\newcommand{\Id}{\mathop{\rm Id}\nolimits}
\newcommand{\defeq}{\buildrel\rm def\over =}
\newtheorem{theorem}{Theorem}
\newtheorem{corollary}{Corollary}
\newcommand{\qedsymbol}{QED}
\newenvironment{proof}{{\em Proof}.}{\qedsymbol}

\begin{document}

\title{Entanglement purification via separable superoperators}

\author{Eric M. Rains}
 
\address{AT\&T Research, Room C290, 180 Park Ave.,
Florham Park, NJ 07932-0971, USA}

\date{February 16, 1998}
\maketitle


\begin{abstract}

One of the fundamental concepts of quantum information theory is that of
entanglement purification; that is, the transformation of a partially
entangled state into a smaller-dimensional, more completely entangled
state.  Of particular interest are protocols for entanglement purification
(EPPs) that alternate purely local operations with one- or two-way
classical communication.  In the present work, we consider a more general,
but simpler, class of transformations, called separable superoperators.
Since every EPP is a separable superoperator, bounds on separable
superoperators apply as well to EPPs; we use this fact to give a new upper
bound on the rate of EPPs on Bell-diagonal states, and thus on the capacity
of Bell-diagonal channels.

\end{abstract}


\narrowtext



\renewcommand{\arraystretch}{.33}

One of the central questions in quantum information theory is that of
determining the capacity of quantum channels; that is, the transmission
rate below which noiseless transmission of entanglement is possible.  In
\cite{BennettEtAl}, Bennett {\em et. al.} reduce this problem to that of
{\em entanglement purification}, the production of maximally entangled
states from non-maximally entangled states.  In particular, Bennett {\em
et. al.}  define two measures of {\em distillable entanglement} for a given
state $\rho$: $D_1(\rho)$, the rate at which singlet states can be produced
from a stream of systems in state $\rho$ using local operations together
with one-way classical communication, and $D_2(\rho)$, the rate when
two-way classical communication is allowed.  While $D_2(\rho)$ is clearly
the maximum that can be physically achieved, the set of allowed
transformations is extremely complicated.  For this reason, we will
introduce a third measure $D_*(\rho)$, or {\em separably distillable
entanglement}, which, while it allows unphysical operations, is more
amenable to analysis.  In particular, we will derive an upper bound on
$D_*(\rho)$, which then immediately gives a bound on $D_2(\rho)$.

Let $\rho$ be a mixed state on a bipartite Hilbert space $V\otimes V$.  We
define the {\em entanglement fidelity} of $\rho$ as
\begin{equation}
F(\rho) = \phi^+(V)^\dagger \rho \phi^+(V),
\end{equation}
where $\phi^+(V)$ is the maximally entangled state
\begin{equation}
{1\over \sqrt{\dim(V)}}
\sum_{0\le i<\dim(V)}
\ket{i}\otimes \ket{i}.
\end{equation}
Note that $\phi^+(V)$ does depend on the basis chosen for $V$, but only up
to local unitary operations.  For any set ${\cal S}$ of physical
transformations, the distillable entanglement $D_{\cal S}(\rho)$ is defined
as the largest number such that there exists a sequence of transformations
$P_i\in {\cal S}$, with $P_i$ mapping states on $V^{\otimes n_i}$ to states
on $W_i$, such that
\begin{equation}
\lim_{i\to\infty} F(P_i(\rho)) = 1
\end{equation}
and
\begin{equation}
\lim_{i\to\infty} {\log_2 \dim W_i\over n_i} = D_{\cal S}(\rho).
\end{equation}
In other words, $D_{\cal S}(\rho)$ is the rate at which entanglement can
be distilled from a stream of systems in the state $\rho$, using only
transformations from ${\cal S}$.  The 1-locally distillable entanglement
$D_1(\rho)$ corresponds to the case when ${\cal S}$ consists of 1-local
operations, that is local operations together with one-way classical
communication, and analogously for $D_2(\rho)$.

For the new measure $D_*(\rho)$, we take the set ${\cal S}$ to be the set
of {\it separable superoperators}.  Recall that if $V$ and $W$ are
(finite-dimensional) Hilbert spaces, a {\em superoperator} (more correctly,
a completely positive trace-preserving map) ${\cal A}$ from $V$ to $W$ is a
linear transformation from operators on $V$ to operators on $W$, such that
${\cal A}\otimes \Id(V')$ maps density operators on $V\otimes V'$ to density
operators on $W\otimes V'$, for any $V'$.  Clearly, any physical
transformation must be a superoperator; moreover, it can be shown
\cite{Superoperator} that any superoperator is realizable via unitary
operations and partial traces.  Moreover, a superoperator can always be
written in the form
\begin{equation}
\rho\mapsto \sum_i A_i \rho A_i^\dagger,
\label{superop}
\end{equation}
where the $A_i$ are linear transformations from $V$ to $W$ such that
\begin{equation}
\sum_i A_i^\dagger A_i = \Id(V),
\end{equation}
although such representation is by no means unique.
If $V=V_1\otimes V_2$ and $W=W_1\otimes W_2$, a {\it separable}
superoperator is one which has a representation of the form (\ref{superop}),
in which each $A_i=A^{(1)}_i\otimes A^{(2)}_i$, with
$A^{(j)}_i$ a linear transformation from $V_j$ to $W_j$.
Clearly, the space of separable superoperators contains that of 1-local
superoperators.  Since the space of separable superoperators is closed
under multiplication, and symmetric under exchanging of $V_1$ and $V_2$,
it follows that every 2-local superoperator is separable.

Remark.  Separable superoperators were implicitly introduced in
\cite{Vedraletal}.  There, however, it was implied that the space of
separable superoperators is identical to the space of 2-local
superoperators, which is certainly not obviously true
(and, indeed, can be shown to be false \cite{sepsup}).  It is quite
possible, therefore, that $D_*(\rho)$ is strictly greater than $D_2(\rho)$
for some states $\rho$.

It will be helpful to observe that there is a natural correspondence
between linear transformations from $V$ to $W$ and vectors in $V\otimes W$.
If $\ket{i}$ is an orthonormal basis of $V$,
and $A$ is a linear transformation from $V$ to $W$,
then we define a vector
\begin{equation}
\ket{A} = \sum_i \ket{i} \otimes A\ket{i} =
\sqrt{\dim(V)} (\Id(V)\otimes A) \phi^+(V).
\end{equation}
We have the following identities:
\begin{eqnarray}
\ket{A}&=&\sqrt{\dim(W)} (A^t\otimes\Id(W)) \phi^+(W)\\
\bra{A}\ket{B}&=&\Tr(A^\dagger B)\\
\Tr_V(\ket{B}\bra{A})&=& B A^\dagger\\
\Tr_W(\ket{B}\bra{A})&=& (A^\dagger B)^t
\end{eqnarray}
In particular, it follows that for a superoperator ${\cal P}$,
\begin{equation}
\Tr_W(\sum_i \ket{P_i}\bra{P_i})
=
\Id(V).
\label{superopvec}
\end{equation}

Fix a separable superoperator ${\cal P}$ from $V\otimes V$ to $W\otimes W$,
where $V$ is an $n$-qubit Hilbert space, and $W$ has dimension $K$; to
be explicit, take
\begin{equation}
{\cal P}(\rho) = \sum_i (P^{(1)}_i\otimes P^{(2)}_i) \rho (P^{(1)}_i\otimes
P^{(2)}_i)^\dagger.
\end{equation}
To any state $\rho$ on $V\otimes V$, we can associate a fidelity $F_{\cal
P}(\rho)$ between $0$ and $1$, namely the fidelity of ${\cal P}(\rho)$.
Consider, in particular, the case in which $\rho$ is the pure state
\begin{equation}
\rho(U)=2^{-n} \ket{U^t}\bra{U^t},
\end{equation}
where $U$ is an arbitrary unitary operation.  Then
\begin{eqnarray}
&&F_{\cal P}(U)\defeq F_{\cal P}(\rho(U))\nonumber\\
&&\quad={1\over 2^n K} \sum_i
\left|\bra{\Id(W)} (P^{(1)}_iU \otimes P^{(2)}_i) \ket{\Id(V)}\right|^2\\
&&\quad={1\over 2^n K}\sum_i
\left|\bra{\overline{P^{(1)}_i}} (U\otimes \Id(W)) \ket{P^{(2)}_i}\right|^2\\
&&\quad=
{1\over 2^n K}\sum_i \Tr\Bigl(\overline{\rho^{(1)}_i}(U \otimes \Id(W))
 \rho^{(2)}_i
(U^\dagger \otimes \Id(W))\Bigr),\nonumber\\
&&
\end{eqnarray}
where
\begin{equation}
\rho^{(1)}_i=\ket{P^{(1)}_i}\bra{P^{(1)}_i},
\end{equation}
and similarly for $\rho^{(2)}_i$.
In particular, $\overline{\rho^{(1)}_i}$ and $\rho^{(2)}_i$ are
positive semi-definite Hermitian operators on $V\otimes W$.
	
If $\rho$ is a mixture of states of the form $\rho(U)$, then $F_{\cal
P}(\rho)$ can be written as a linear combination of the relevant $F_{\cal
P}(U)$s.  Of particular interest is the case of the ``depolarizing'' qubit
state; that is, the state $\Delta(\epsilon)$ with density matrix
\begin{equation}
{\epsilon\over 4}\Id+
{1-\epsilon \over 2} (\ket{00}+\ket{11})(\bra{00}+\bra{11}),
\end{equation}
on a two-state Hilbert space.  We can also write
\begin{equation}
\Delta(\epsilon) = 
f \rho(1) + {1-f\over 3} (\rho(\sigma_x)+\rho(\sigma_y)+\rho(\sigma_z)),
\end{equation}
with $f=1-(3/4)\epsilon=F(\Delta(\epsilon))$.  We can then write:
\begin{equation}
F_{\cal{P}}(\Delta(\epsilon)^{\otimes n})
=
\sum_j f^{n-j} ({1-f\over 3})^j
\sum_{\wt(E)=j}
F_{\cal P}(E),
\end{equation}
where $E$ ranges over the set ${\cal E}$ of tensor products of matrices
from the set $\{1,\sigma_x,\sigma_y,\sigma_z\}$, and $\wt(E)$ is
the number of components in the tensor product not equal to the identity.
The quantity
\begin{equation}
B_j({\cal P}) = \sum_{\wt(E)=j} F_{\cal P}(E)
\end{equation}
has a form very similar to that of the weight enumerators studied in
\cite{ShorLaflamme} and \cite{UnitaryPap}; this suggests that we should
consider the quantity
\begin{equation}
B'_S({\cal P}) = {1\over 2^n K} \sum_i
\Tr(\Tr_S(\overline{\rho^{(1)}_i})\Tr_S(\rho^{(2)}_i)),
\end{equation}
where $\Tr_S(\rho)$ is the partial trace of $\rho$ with respects to the
qubits of $V$ indexed by $S$, as well as $W$ if $0\in S$.
We can then define a polynomial
\begin{equation}
B'(u,v,x,y)
=
\sum_i x^{n-i} y^i 
 \!\!\!\!\!\!\! \sum_{S\subset \{1,2,\ldots n\}\atop |S|=i}\!\!\!\!\!\!\!
   \big( u B'_S({\cal P}) + v B'_{\{0\}\cup S}({\cal P})\big).
\end{equation}
The arguments in \cite{UnitaryPap} tell us that
\begin{equation}
B'(1,0,x-y,2y)=
B(x,y)=\sum_i B_i({\cal P}) x^{n-i} y^i.
\end{equation}
On the other hand,
\begin{eqnarray}
&&B'_{\{0\}\cup S}({\cal P})
\nonumber\\
&&\quad=
{1\over 2^n K}
\sum_i \Tr\left(\Tr_S(\Tr_W(\overline{\rho^{(1)}_i}))
\Tr_S(\Tr_W(\rho^{(2)}_i))\right)\\
&&\quad=
{1\over 2^n K}
\sum_i \bra{\Id}
\Tr_{S\times S}\Bigl(\Tr_{W\otimes
W}(\rho^{(1)}_i\otimes \rho^{(2)}_i)\Bigr)
\ket{\Id}
\end{eqnarray}
Since ${\cal P}$ is a superoperator, (\ref{superopvec}) tells us that
\begin{equation}
\sum_i
\Tr_{W\otimes W}(\rho^{(1)}_i\otimes \rho^{(2)}_i)
=
\Id(V\otimes V).
\end{equation}
It follows that
\begin{equation}
B'_{\{0\}\cup S}({\cal P})
=
2^{|S|}/K,
\end{equation}
and thus that
\begin{equation}
B'(u,v,x,y) = u B(x+y/2,y/2) + v (x+2y)^n/K.
\label{B'eq}
\end{equation}

Since each $\overline{\rho^{(1)}_i}$ and $\rho^{(2)}_i$ is positive
semi-definite, the theory of weight enumerators \cite{UnitaryPap} tells us
that the polynomials $B'(u-v,K v,x-y,2y)$ and $B'(v-u,u+v,y-x,x+y)$ have
nonnegative coefficients; note that the latter is the analogue of the
``shadow'' enumerator, which was shown to be nonnegative in
\cite{Invariants}.  These polynomials can be written in terms of $B(x,y)$,
using (\ref{B'eq}):
\begin{eqnarray}
&&B'(u-v,Kv,x-y,2y)\nonumber\\
&&\quad =
u B(x,y) + v ((x+3y)^n-B(x,y)),\\
&&B'(v-u,u+v,y-x,x+y)\nonumber\\
&&\quad =
\phantom{+} u ({1\over K} (x+3y)^n-S(x,y)) \nonumber\\
&&\phantom{\quad = } + v ({1\over K} (x+3y)^n+S(x,y)),
\end{eqnarray}
where
\begin{equation}
S(x,y)=B({3y-x\over 2},{x+y\over 2}).
\end{equation}
Since both of those polynomials have nonnegative coefficients, we can
conclude that the four polynomials
\begin{eqnarray}
&B(x,y),\ (x+3y)^n-B(x,y),&\nonumber\\
&{1\over K}(x+3y)^n-S(x,y),\ {1\over K}(x+3y)^n+S(x,y),&\nonumber
\end{eqnarray}
each have nonnegative coefficients.
The first two polynomials simply correspond to the fact that
\begin{equation}
0\le F_{\cal P}(E)\le 1
\label{Bineq}
\end{equation}
for all $E$.  The second pair of polynomials roughly say that
\begin{equation}
|S_{\cal P}(E)|\le {1\over K},
\end{equation}
for an appropriate definition of $S_{\cal P}(E)$; it is not clear
what, if anything, this corresponds to physically.

We can now begin to obtain bounds on $D_*$:

\begin{theorem}
Let ${\cal P}$ be a separable superoperator from $V\otimes V$ to $W\otimes
W$, where $V$ is an $n$-qubit Hilbert space, and $W$ is a $K$-dimensional
Hilbert space.  Then for any $f\le {1\over 2}$,
\begin{equation}
F_{\cal P}(f)\defeq F_{\cal P}(\Delta((4/3)(1-f)))\le {1\over K}.
\label{shadowbound}
\end{equation}
In particular, $D_*(f)=0$.
\end{theorem}

\begin{proof}
We have:
\begin{eqnarray}
F_{\cal P}(f)&=& B(f,{1-f\over 3})\\
             &=& S({1\over 2}-f,{1\over 6}+{f\over 3}).
\end{eqnarray}
Now, the coefficients of $(x+3y)^n/K-S(x,y)$ are nonnegative, so, for
any specific numbers $x,y\ge 0$,
\begin{equation}
(x+3y)^n/K\ge S(x,y).
\end{equation}
In particular, this is true for $x=(1/2)-f$ and $y=(1/6)+(f/3)$; the
result follows immediately.
\end{proof}

In particular, we obtain the known fact that distillation is impossible for
$f\le {1\over 2}$.  Moreover, we obtain the following:

\begin{corollary}
If $\rho$ is a separable state on $W\otimes W$, where $W$ has dimension
$K$, then $\rho$ has fidelity at most $1/K$.  In particular, for any
separable state $\chi$, and any separable superoperator ${\cal P}$,
$F_{\cal P}(\chi)\le 1/K$.
\end{corollary}

\begin{proof}
Suppose, on the other hand, that $\rho$ had fidelity greater than $1/K$.
Since $\rho$ is separable, we could then produce a $K\times K$-dimensional
bipartite state of fidelity greater than $1/K$ from any input state, using
only local operations and classical communication.  But this contradicts
the bound (\ref{shadowbound}).  The second statement follows from the fact
that the image of a separable state under a separable superoperator is
separable.
\end{proof}

It should also be noted that (\ref{shadowbound}) is tight, since a
uniformly distributed ensemble of states $\psi\otimes\psi$ is certainly
separable, and is easily shown to have fidelity $1/K$; the argument of the
corollary then applies in reverse to construct a separable superoperator of
fidelity $1/K$.

So far, we have not used the first two constraints.  It turns out that
these can be used to control how much the output fidelity of a given
superoperator can vary as the input fidelity changes.  In particular,
we will be able to establish, for each rate, a neighborhood of $f={1\over
2}$ for which the output fidelity must still tend to 0.

\begin{theorem}
Let $f\ge {1\over 2}$.  The separably distillable entanglement $D_*(f)$ of
the depolarizing state $\Delta(\epsilon)$ with fidelity $f$, satisfies the
bound
\begin{eqnarray}
D_*(f)&\le& 1-H_2(f)\\
      &=  & 1+f\log_2(f)+(1-f)\log_2(1-f).
\end{eqnarray}
Indeed, any family of separable superoperators of rate greater than
$1-H_2(f)$ must have output fidelity tending to 0.
\end{theorem}

\begin{proof}
Suppose the theorem were false.  Then there would exist a sequence
of separable superoperators ${\cal P}_i$, producing a $K_i\times K_i$
bipartite state from $n_i+n_i$ qubits such that $F_{{\cal P}_i}(f)$
did not tend to 0, and such that $\log_2(K_i)/n_i$ tended to a limit
strictly greater than $1-H_2(f)$.

Consider $B(f,{1-f\over 3})$.  For a fixed value of $B(f,{1-f\over 3})$,
the lowest possible value of $B(1/2,1/6)$ (ignoring all other constraints)
is attained when the weight of the $B_i$ is concentrated at low $i$; these
are the coefficients for which $f^{n-i}((1-f)/3)^i$ is decreased the most
when $f$ is replaced by $1/2$.
In that case, we have:
\begin{equation}
B(f,{1-f\over 3})\simeq \sum_{0\le i<j} {n\choose i} f^{n-i} (1-f)^i
\end{equation}
for some $j$.  In order for this not to tend to 0 as $n$ increases, we must
have $j\gtrsim n(1-f)$.  But then
\begin{equation}
B({1\over 2},{1\over 6})\simeq 2^{-n} \sum_{0\le i<j} {n\choose i},
\end{equation}
so
\begin{equation}
B({1\over 2},{1\over 6})\simeq 2^{n (H_2(j/n)-1)}\gtrsim 2^{n (H_2(f)-1)}.
\end{equation}
On the other hand, by (\ref{shadowbound}), we know $B(1/2,1/6)\le 1/K$.
But then
\begin{equation}
\log_2(K)/n\lesssim 1-H_2(f).
\end{equation}
\end{proof}

This bound is plotted in Figure 1, as well as the weaker bound
\begin{equation}
D_2(f)\le E(f)=H_2({1\over 2}+\sqrt{f(1-f)})
\end{equation}
from \cite{BennettEtAl}.  In particular, note that the new bound is
strictly stronger than the old bound (``entanglement of formation'') for
$1/2<f<1$.  Since every 1-local operator is separable, we also get the
bound $D_1(f)\le 1-H_2(f)$, which actually improves on the best known upper
bounds, for some range of $f$.  For $1/2\le f\le 3/4$, it is known that
1-local operators cannot achieve fidelity close to 1 at any positive rate;
if this could be strengthened to more precise bounds on fidelity, the above
technique would then provide bounds on $D_1(f)$ for $f\ge 3/4$.

The above argument can be extended to arbitrary
Bell-diagonal states; to bound $D_*(\chi)$ where $\chi$ is Bell-diagonal
with eigenvalues $\beta_0\ge\beta_1\ge \beta_2\ge \beta_3$, with
$\beta_0\ge 1/2$, simply compare $\chi$ to the separable Bell-diagonal
state $\chi_0$ with eigenvalues $1/2$, $\beta_1/(2-2\beta_0)$,
$\beta_2/(2-2\beta_0)$, and $\beta_3/(2-2\beta_0)$.  The separability of
$\chi_0$ implies, by corollary 1, that $F_{\cal P}(\chi_0)\le 1/K$; but
then (\ref{Bineq}) allows us to deduce that $F_{\cal P}(\chi)$ tends to 0
unless
\begin{equation}
{\log_2(K)\over n}
\lesssim
1-H_2(\beta_0).
\label{mybound2}
\end{equation}
In other words, $D_*(\chi)\le 1-H_2(\beta_0)$.  Note that this bound is
tight in the case $\beta_2=\beta_3=0$; in this case, the noise is purely
classical in nature, and can be corrected using classical codes.

Vedral and Plenio \cite{VedralPlenio} have independently proved
(assuming a certain additivity conjecture) a more general bound on $D_*$,
which apparently agrees with (\ref{mybound2}) on Bell-diagonal states.

The author would like to thank Andr\'e Berthiaume for helpful comments,
as well as David DiVincenzo, Peter Shor, and John Smolin for helpful
conversations about separable superoperators.

\begin{figure}
\centerline{\epsfbox{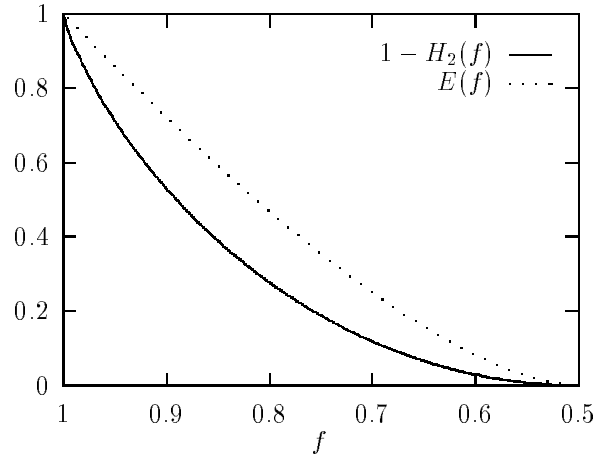}}
\caption{Bounds on $D_2(f)$}
\end{figure}


\begin{thebibliography}{2}

\bibitem{BennettEtAl} C. H. Bennett, D. DiVincenzo, J. A. Smolin, and
W. K. Wootters, Mixed state entanglement and quantum error correction,
Phys. Rev. A 54, 3824 (1996), also LANL e-print quant-ph/9604024.

\bibitem{Superoperator} B. W. Schumacher, Sending entanglement through
noisy channels, Phys. Rev. A 54, 2614 (1996), also LANL e-print
quant-ph/9604023.

\bibitem{Vedraletal}
V. Vedral, M. B. Plenio, M. A. Rippin, and P. L. Knight,
Quantifying entanglement, Phys. Rev. Lett., {\bf 78} (1997) 2275--2279,
also LANL e-print quant-ph/9702027.

\bibitem{sepsup}
C. H. Bennett, D. P. DiVincenzo, C. Fuchs, P. H\o yer, T. Mor,
E. M. Rains, P. W. Shor, and J. Smolin,
Quantum nonlocality without entanglement, manuscript in preparation.

\bibitem{ShorLaflamme}
P. W. Shor and R. A. Laflamme, Quantum analog of the MacWilliams identities
in classical coding theory, Phys. Rev. Lett., {\bf 78} (1997) 1600--1602, 
also LANL e-print quant-ph/9610040.

\bibitem{UnitaryPap}
E. M. Rains, Quantum weight enumerators,
LANL e-print quant-ph/9612015.

\bibitem{Invariants}
E. M. Rains, Polynomial invariants of quantum codes,
LANL e-print quant-ph/9704042.

\bibitem{VedralPlenio}
V. Vedral and M. B. Plenio,
Entanglement measures and purification procedures,
LANL e-print quant-ph/9707035.

\end{thebibliography}
\end{document}